\pdfoutput=1
\documentclass[referee]{raa}

\usepackage{graphicx,times}
\usepackage{natbib}
\usepackage{amssymb,amsmath}
\bibpunct{(}{)}{;}{a}{}{,}

\usepackage{threeparttable}

\usepackage{CJKutf8}
\usepackage[colorlinks,citecolor=cyan]{hyperref}

\begin{document}

  \title{Constraining the orbital inclination and companion properties of three black widow pulsars detected by FAST}

   \volnopage{Vol.0 (20xx) No.0, 000--000}
   \setcounter{page}{1}

   \author{Ze-Xin Du \begin{CJK*}{UTF8}{gbsn} (杜泽昕) \end{CJK*}
      \inst{1}
   \and Yun-Wei Yu \begin{CJK*}{UTF8}{gbsn} (俞云伟) \end{CJK*}
      \inst{1, 2}
   \and A-Ming Chen
      \inst{3}
    \and Shuang-Qiang Wang \begin{CJK*}{UTF8}{gbsn} (王双强) \end{CJK*}
      \inst{4}
    \and Xia Zhou \begin{CJK*}{UTF8}{gbsn} (周霞) \end{CJK*}
      \inst{4,5,6}
    \and Xiao-Ping Zheng \begin{CJK*}{UTF8}{gbsn} (郑小平) \end{CJK*}
      \inst{1,2}
   }

   \institute{Institute of Astrophysics, Central China Normal University, Wuhan 430079, China; {\it yuyw@mail.ccnu.edu.cn}\\
        \and
             Key Laboratory of Quark and Lepton Physics (Central China Normal University), Ministry of Education, Wuhan 430079, China\\
        \and
             Tsung-Dao Lee Institute, Shanghai Jiao Tong University, Shanghai 201210, China; {\it chensm@mails.ccnu.edu.cn}\\
        \and 
             Xinjiang Astronomical Observatory, Chinese Academy of Sciences, Urumqi 830011, China\\
        \and
             Key Laboratory of Radio Astronomy, Chinese Academy of Sciences, Nanjing 210008, China\\
        \and 
            Xinjiang Key Laboratory of Radio Astrophysics, Urumqi 830011, China\\
\vs\no
   {\small Received 2023 month day; accepted 2023 month day}}

\abstract{Black widows (BWs) are millisecond pulsars ablating their companion stars. 
The out-flowing material from the companion can block the radio emission of the pulsar, resulting in eclipses. 
In this paper, we construct a model for the radio eclipse by calculating the geometry of the bow shock between the winds of the pulsar and companion, where the shock shapes the eclipsing medium but had not been described in detail in previous works. The model is further used to explain the variations of the flux density and dispersion measure (DM) of three BW pulsars (i.e., PSR B1957$+$20, J2055$+$3829, and J2051$-$0827) detected by the Five-hundred-meter Aperture Spherical radio Telescope (FAST). 
Consequently, we constrained the parameters of the three BW systems such as the inclination angles and true anomalies of the observer as well as the mass-loss rates and wind velocity of the companion stars.
With the help of these constraints, it is expected that magnetic fields of companion stars and even masses of pulsars could further be determined as some extra observation can be achieved in the future.
\keywords{pulsars: individual: PSR B1957$+$20, PSR J2055$+$3829, PSR J2051$-$0827 -- binaries: eclipsing -- Stars: winds, outflows}
}

   \authorrunning{Du et al. 2023 }
   \titlerunning{Constraining the properties of companions of BWs}

   \maketitle

\section{Introduction}
\label{sect:intro}
Millisecond pulsars (MSPs) are rapidly rotating neutron stars, with a large number of them being discovered in binary systems. It is widely believed that MSPs are spun up via angular momentum transfer from their companion star through accretion processes \citep{Bhattacharya_1996}. 
Redbacks (RBs) and black widows (BWs), widely known as spider pulsars, are sub-populations of MSPs that can evaporate their donors persistently by their powerful winds. Masses of RB companion stars are usually within the range of $\sim (0.2-0.4) M_{\odot}$, while BW companions with a range of $\sim (0.02-0.05) M_{\odot}$ \citep{Roberts_2011, Polzin_2019}. The optical observations of RB and BW binaries confirmed that the surface of their companion stars could be strongly irradiated by the energetic pulsar winds \citep{Khechinashvili_2000}.
Furthermore, observations at higher energies also discovered the X-ray emission arising from the interaction of the pulsar wind with the evaporation material \citep{Kluzniak_1988, Phinney_1988, Wu_2012, Huang_2012}.

According to observations, the eclipse regions of spider pulsar binaries are typically larger than the Roche lobes of the companion stars, and therefore the eclipse is not solely caused by the companion star's obscuration, but probably due to the absorption of the radio pulse signal by the surrounding plasma around the companion star, especially during the beginning and ending stages of the eclipse. Since the low-mass companion stars are surrounded by a diffuse evaporation halo, the radio emission from the MSPs can in principle be eclipsed by the evaporation halo as it appears on the line of sight (LOS) of observers near inferior conjunctions \citep{Guillemot_2019}. Such eclipses can happen as long as the orbital inclination of the binary is high enough. 
The first discovered eclipsing BW in radio observation is PSR B1957$+$20 \citep{Fruchter_1988}. About 10\% of its orbital phase was found to be eclipsed, which cannot be covered by the range of the Roche lobe of its companion and thus the ablated material must replenish the eclipse medium to be beyond the Roche lobe \citep{Ray_2017, Polzin_2019}. Therefore, it is expected that the study of the MSP eclipses can help to probe the property of the evaporation material and, furthermore, of the companion stars. Meanwhile, a constraint on the orbital parameters of the binaries could also be obtained. 

The physical processes of eclipsing pulsars had been investigated thoroughly by \cite{Thompaon_1994}, by confronting the observational results of PSR B1957$+$20 and PSR J1748$-$2446A. The correlation between the eclipse duration and the frequency of PSR J2215$+$5135 was studied by \cite{Broderick_2016}, who suggested that the primary effect of the eclipsing medium is to absorb the radio emission rather than scattering it. \cite{Polzin_2019} reported the radio observations of PSR J2051$-$0827 in the frequency range of 110$-$4032 MHz and found that the scattering and/or cyclotron absorption provides the most promising eclipse mechanism. 
\cite{Kudale_2020} further suggested the absorption is primarily due to the cyclotron-synchrotron process, by explaining the excess dispersion, scattering, and absorption of PSR J1227$-$4853.
In any case, it is worth mentioning that cyclotron absorption requires a strong magnetic field in the eclipse medium. 

In the last decade, many eclipsing pulsar binaries have been discovered, which allows us to study the eclipse mechanisms in more detail \citep{Bhattacharyya_2013, Guillemot_2019, Nieder_2020}. In particular, the Five-hundred-meter Aperture Spherical radio Telescope (FAST) has implemented plenty of radio observations for several BW pulsars such as PSR B1957$+$20, PSR J2055$+$3829, and PSR J2051$-$0827, including long-term measurements of their flux density and dispersion measure \citep[DM;][]{Wang_2023}. Furthermore, rotation measure (RM) variation was found in PSR J2051$-$0827. Therefore, at the current stage, it is necessary to combine these observational results with the eclipse processes to constrain the physical properties of binary systems. 

In the next section, we introduce an eclipse model by taking into account the shock interaction between the winds of the pulsar and companion star as well as the consequent geometry of the shock.
Then, we apply the model to fit the light curves and DM variations of the three FAST BWs in Section 3. The implications of the parameter constraints for the binary systems are discussed in Section 4. 
Finally, a conclusion is given in Section 5.

\section{The model}
\label{sect:Model}

\begin{figure*}[h!]
    \centering
    \includegraphics[scale = 0.35]{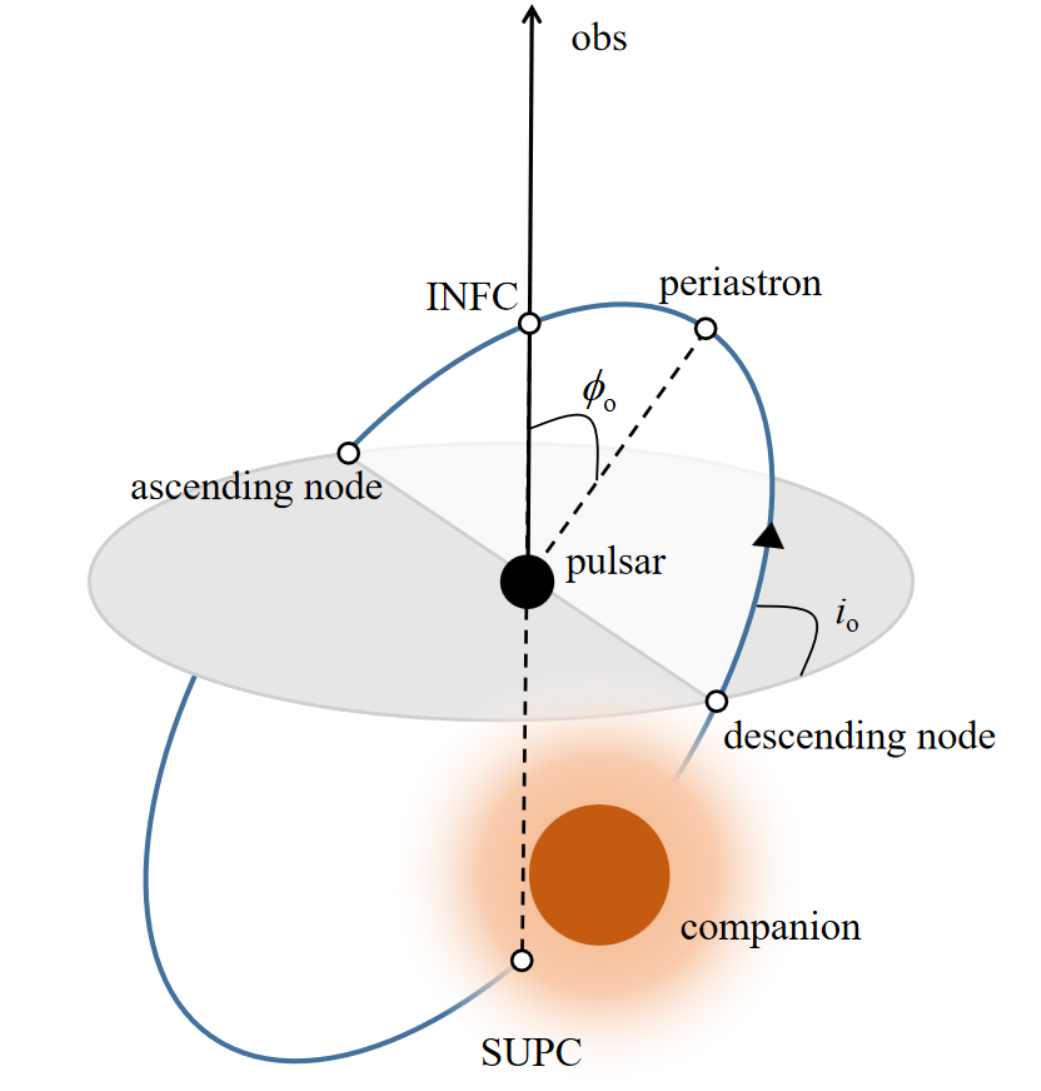}
    \includegraphics[scale = 0.4]{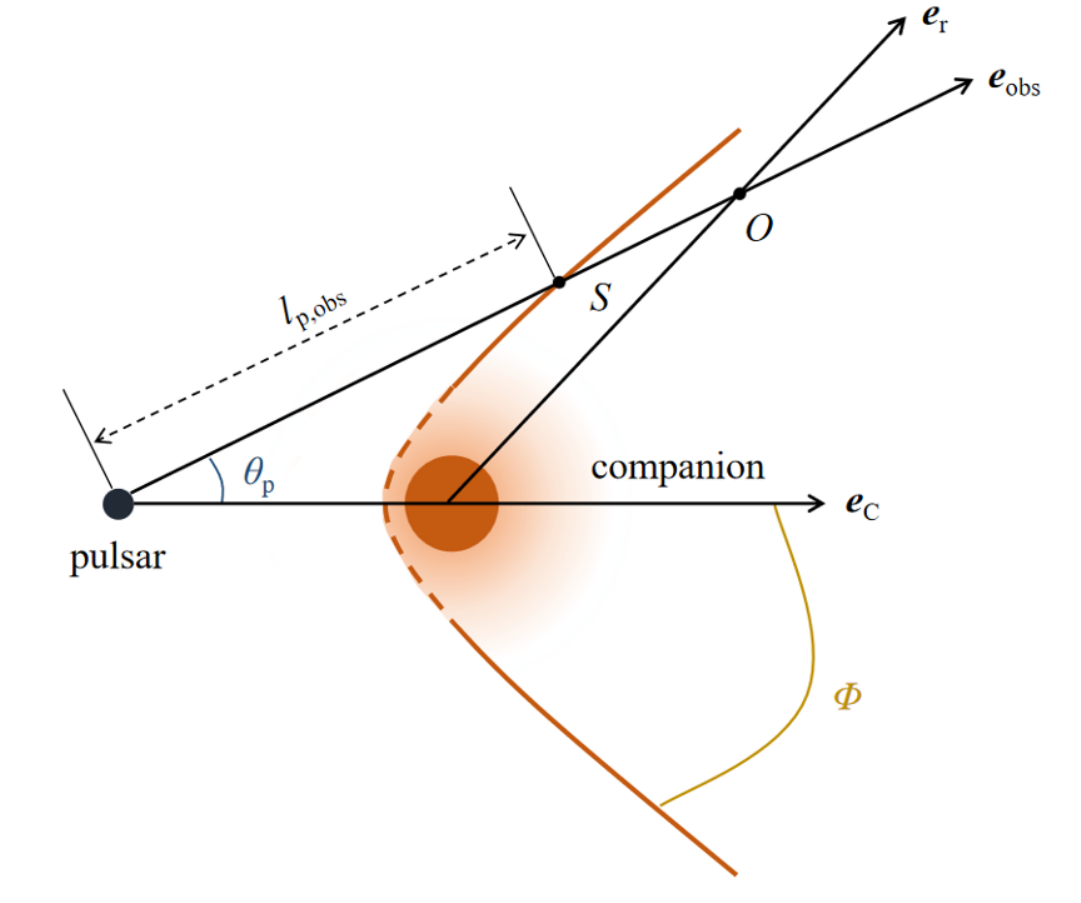}
    \caption{\textit{Left:} Illustration for the general case of the orbital inclination $(i_{\mathrm{o}})$ and the true anomaly of the LOS $(\phi_{\mathrm{o}})$ in spider pulsar binary systems. \textit{Right:} Description of bow shock structure in spider binary systems of the companion star.}
    \label{Fig.1}
\end{figure*}

The geometry of the eclipse medium is determined by the shock surface between the pulsar wind and stellar wind as well as a magnetosphere \citep{Romani_2016, Wadiasingh_2017}. Recently, \cite{Chen_2021} investigated the shock interaction of a pulsar with an O/B star in some high-mass gamma-ray binaries and estimated the radio emission window of these pulsars. 
The primary difference between the spider pulsar binaries from the gamma-ray binaries is that, in the latter case, the intense wind from the O/B star can nearly enclose the pulsar, whereas in the former case, the stellar wind can only blow a conical region. The geometry of the bow shock due to the interaction of the spider pulsar wind with the companion wind is depicted in Fig. \ref{Fig.1}, which is determined by the mechanical balance between the two shocked wind regions. The most crucial parameter is the momentum rate ratio between the winds as:
\begin{equation}
    \eta = \frac{L_{\mathrm{sd}} / c}{\dot{M_{\mathrm{C}}} \, v_{\mathrm{w}}} \label{6},
\end{equation}
where $L_{\mathrm{sd}}$ is the spin-down power of the pulsar, $c$ is the speed of light, $\dot{M_{\mathrm{C}}}$ is the mass loss rate of the companion star, and $v_{\mathrm{w}}$ is the terminal speed of this wind.

Following \cite{Chen_2021}, when the line of sight (LOS) of observers can intersect with the bow shock, the radio emission from the pulsar would be suppressed as $F_{\nu}(\nu) = F_{\nu,0} \cdot e^{- \tau(\nu)}$,
where the optical depth of the companion wind on the LOS is considered to be dominated by the free-free absorption (FFA) process as:
\begin{equation}
    \tau (\nu) = \int_{l_{\mathrm{p,obs}}}^{\infty} \alpha(\nu;T,n_{\rm e},n_{\rm i}) d l \label{7},
\end{equation}
where $\alpha(\nu;T,n_{\rm e},n_{\rm i})$ is the FFA coefficient, which is dependent on the temperature ($T$) of the medium and the density of the electrons ($n_{\rm e}$) and ions ($n_{\rm i}$), and the value of $l_{\mathrm{p, obs}}$ is determined by the shape of the bow shock. Here, the FFA process rather than the cyclotron/synchrotron processes is invoked, because the latter processes usually require a strong magnetic field. The strength of the magnetic field could in principle be measured by detecting the variation of the rotation measure (RM) of the radio emission near the eclipse boundaries \citep{Polzin_2019, Kudale_2020}. For the spider pulsar PSR B1957$+$20, 
\cite{Li_2019} measured its RM values and found no evidence for large-scale magnetic fields over the egress plasma lensing region.
Besides the radio suppression effect, the companion wind can further provide an extra contribution to the dispersion measure (DM) of the pulsar radio emission, which leads to a DM variation as
\begin{align}
    \Delta \mathrm{DM} &= \int_{l_{\mathrm{p,obs}}}^{\infty} n_{\mathrm{e}} \, d l \label{8},
\end{align}
where $n_{\mathrm{e}}$ is the electron number density, which is determined by the hydrogen abundance of the wind and the distance to the centre of the companion star. Specifically, we can relate the electron number density with the ion density by $n_{\mathrm{e}} = n_{\mathrm{i}}\mu_{\mathrm{i}} / \mu_{\mathrm{e}}$, where the typical values of the mean ion molecular and electron weight are taken as $\mu_{\mathrm{i}} \sim 1.29$ and $\mu_e \sim 1.18$, respectively \citep{Zdziarski_2010}. 

In order to complete the above integrals, we write the number density of the companion wind (i.e., the ion density $n_{\rm i}=n_{\rm w}$) as a function of the radius $r$ to the companion as \citep{Waters_1988}:
\begin{equation}
    n_{\mathrm{w}} (r) = n_{\mathrm{w,0}} \left( \frac{r}{r_{\mathrm{\star}}}\right)^{-2} \hfill \label{1},
\end{equation}
where the base density $n_{\mathrm{w,0}}$ at the stellar surface $r_{\mathrm{\star}}$ can be defined as:
\begin{equation}
    n_{\mathrm{w},0} = \frac{\dot{M_{\mathrm{C}}}} {4 \pi {r_{\star}}^2 v_{\mathrm{w}} \mu_{\mathrm{i}} m_{\mathrm{p}}} \label{2},
\end{equation}
where $m_{\mathrm{p}}$ is the mass of protons, $v_{\mathrm{w}}$ is the wind velocity at surface of companion. 
Here the mass-loss of the companion is assumed to be isotropic, which is viable for the region not very close to the companion. In addition, it should be mentioned that when the orbital inclination is high enough (i.e., $i_{\mathrm{o}} > {\pi\over2} - \theta_{\star}$), the companion star can appear on the LOS and obstruct the radio emission completely, where $\theta_{\star}$ is the opening angle of the companion star with respect to the centroid of the binary. 

On the other hand, the temperature of the stellar wind gradually decreases through adiabatic cooling with increasing radial distance from the surface of the companion star. As a result, the temperature distribution can be empirically expressed by a power law as \citep{Kochanek_1993}:
\begin{equation}
    T_{\mathrm{w}} (r) = T_{\mathrm{\star}} \left( \frac{r}{r_{\mathrm{\star}}}\right) ^{- \beta}\equiv A r^{-\beta} \label{Tr}  ,
\end{equation}
where $T_{\mathrm{\star}}$ is the effective temperature of the star at $r_{\mathrm{\star}}$ and $A\equiv T_{\star} r^{ \, \beta}_{\star}$. The value of the index $\beta$ depends on the adiabatic index of the wind gas, which could range within $\sim$ (2/3 $-$ 4/3). In the following calculations, we simply adopt $\beta$ = 2/3. 
Strictly speaking, the stellar surface temperature is actually anisotropic due to the irradiation by the pulsar and thus the temperature $T_{\star}$ involved here can only be treated as an effective one. Nevertheless, it should still be noticed that the temperature directly involved in the calculation is actually the temperature of the wind material far away from the stellar surface, where the anisotropy of the temperature is very likely to be not as significant as on the stellar surface.

By confronting the above model with the radio observations by spider pulsars, we can constrain the model parameters of the binary orbit (the inclination angle $i_{\mathrm{o}}$ and the true anomaly of observer $\phi_{\mathrm{o}}$) and of the companion star ($\dot{M_{\mathrm{C}}}$, $v_{\mathrm{w}}$, and $A$).
 
\begin{table*}[htbp]
   \centering
   \caption{\centering{Parameters and fitting results of PSR B1957$+$20, PSR J2051$-$0827, PSR J2055$+$3829}}
   \begin{threeparttable}
   \begin{tabular}{ccccc}
   \hline
   \hline
   \textbf{Parameters} & \textbf{PSR B1957$+$20} & \textbf{PSR J2051$-$0827} & \textbf{PSR J2055$+$3829}\\
   \hline

   orbital period, $P_{\mathrm{orb}}$ (hr) & $9.2^{ \, (\mathrm{a})}$ & $2.38^{\, (\mathrm{d})}$ & $3.1^{\, (\mathrm{g})}$ \\

   spin-down luminosity, $L_{\mathrm{sd}}$ ($\mathrm{erg \, s^{-1}}$) & $1.6 \times 10^{35 \, (\mathrm{b})}$ & $5.5 \times 10^{33 \, (\mathrm{e})}$ & $4.3 \times 10^{33 \, (\mathrm{g})}$ \\

   eccentricity, $e$ ($10^{-5}$) & $4.0^{\, (\mathrm{a})}$ & $6.2^{\, (\mathrm{f})}$ & $1.4^{\, (\mathrm{g})}$\\

   projected binary separation, $a_{\mathrm{orb}} \mathrm{sin} i_\mathrm{o}$ ($\mathrm{R_{\odot}}$) & $2.7^{\, (\mathrm{c})}$ & $1.02^{\, (\mathrm{f})}$ & $1.2^{\, (\mathrm{g})}$\\
   \hline

   inclination angle of observer, $i_{\mathrm{o}}$ ($\mathrm{^{\circ}}$) & $85.06^{+0.38}_{-0.37}$ & $59.50^{+0.40}_{-0.36}$ & $46.80^{+1.53}_{-1.64}$\\

   true anomaly of observer, $\phi_{\mathrm{o}}$ ($\mathrm{^{\circ}}$) & $94.77^{+0.02}_{-0.02}$ & $81.22^{+0.02}_{-0.03}$ & $100.48^{+0.05}_{-0.05}$\\

   mass-loss rate, $\mathrm{log_{10}} \dot{M_{\mathrm{C}}}$ ($\mathrm{M_{\odot} \, yr^{-1}}$) & $-12.15^{+0.02}_{-0.02}$ & $-11.92^{+0.01}_{-0.01}$ & $-11.64^{+0.04}_{-0.03}$\\

   wind velocity, $v_{\mathrm{w}}$ ($10^{8} \, \mathrm{cm \, s^{-1}}$) & $2.05^{+0.08}_{-0.07}$ & $0.54^{+0.01}_{-0.00}$ & $1.03^{+0.01}_{-0.01}$\\

   temperature coefficient, $A~(10^{10} \, \mathrm{K \, cm^{2/3}})$ & $0.23^{+0.01}_{-0.01}$ & $2.78^{+0.06}_{-0.05}$ & $0.92^{+0.01}_{-0.01}$\\
   \hline
   evaporation efficiency $f~(10^{-2})$ & 0.235 & 3.33 & 4.34 \\
   \hline
   \end{tabular}
   \begin{tablenotes}
   \item \textbf{Reference.} (a) \cite{Fruchter_1988}; (b) \cite{Arzoumanian_1994}; (c) \cite{Lin_2023}; (d) \cite{Stappers_1996_apjl}; (e)
   \cite{Shaifullah_2016}; (f) \cite{Lazaridis_2011}; (g) \cite{Guillemot_2019}
   \end{tablenotes}
   \end{threeparttable}
   \label{T1}
\end{table*}

\section{Fitting the flux and DM variations}
\label{sect:Fitting}

\subsection{PSR B1957$+$20} \label{subsec:PSR B1957+20}

\begin{figure*}[h!]
    \centering
    \includegraphics[scale = 0.4]{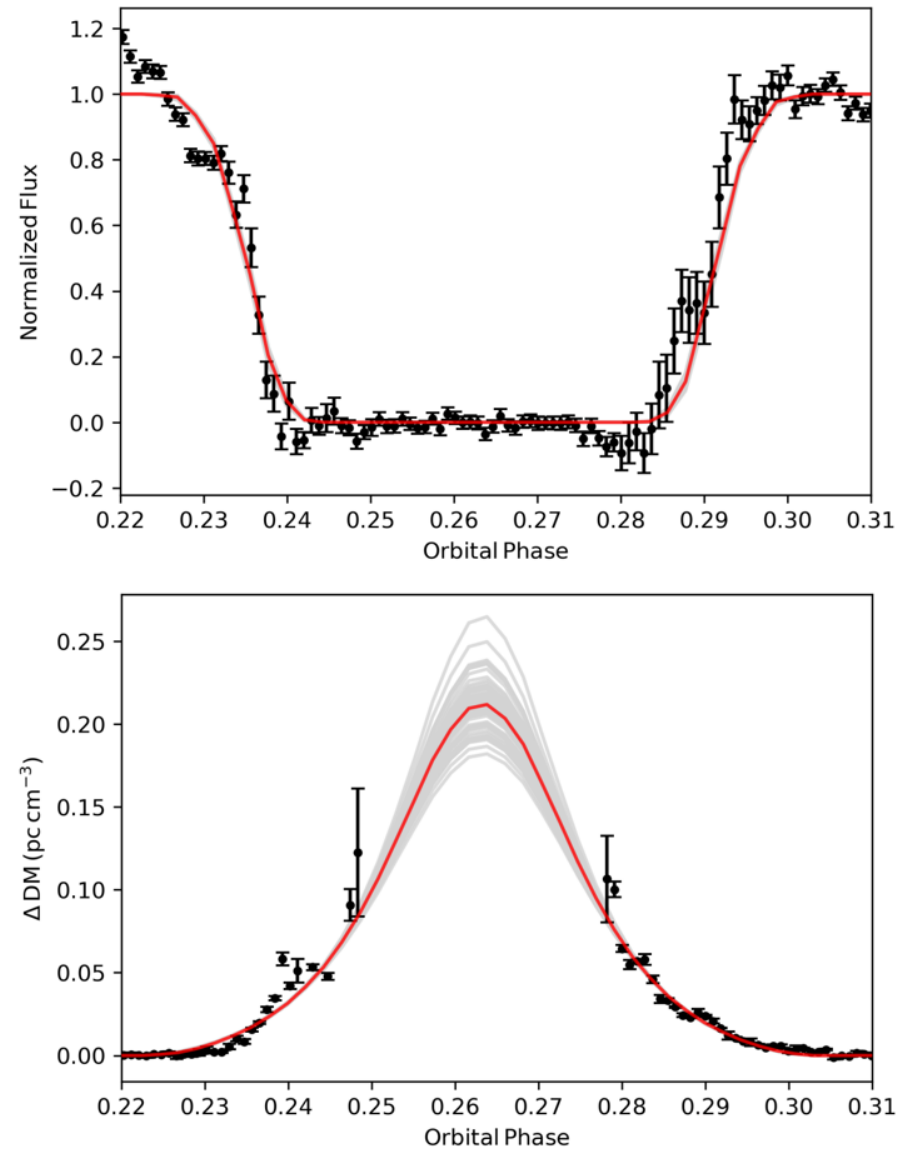}
    \includegraphics[scale = 0.25]{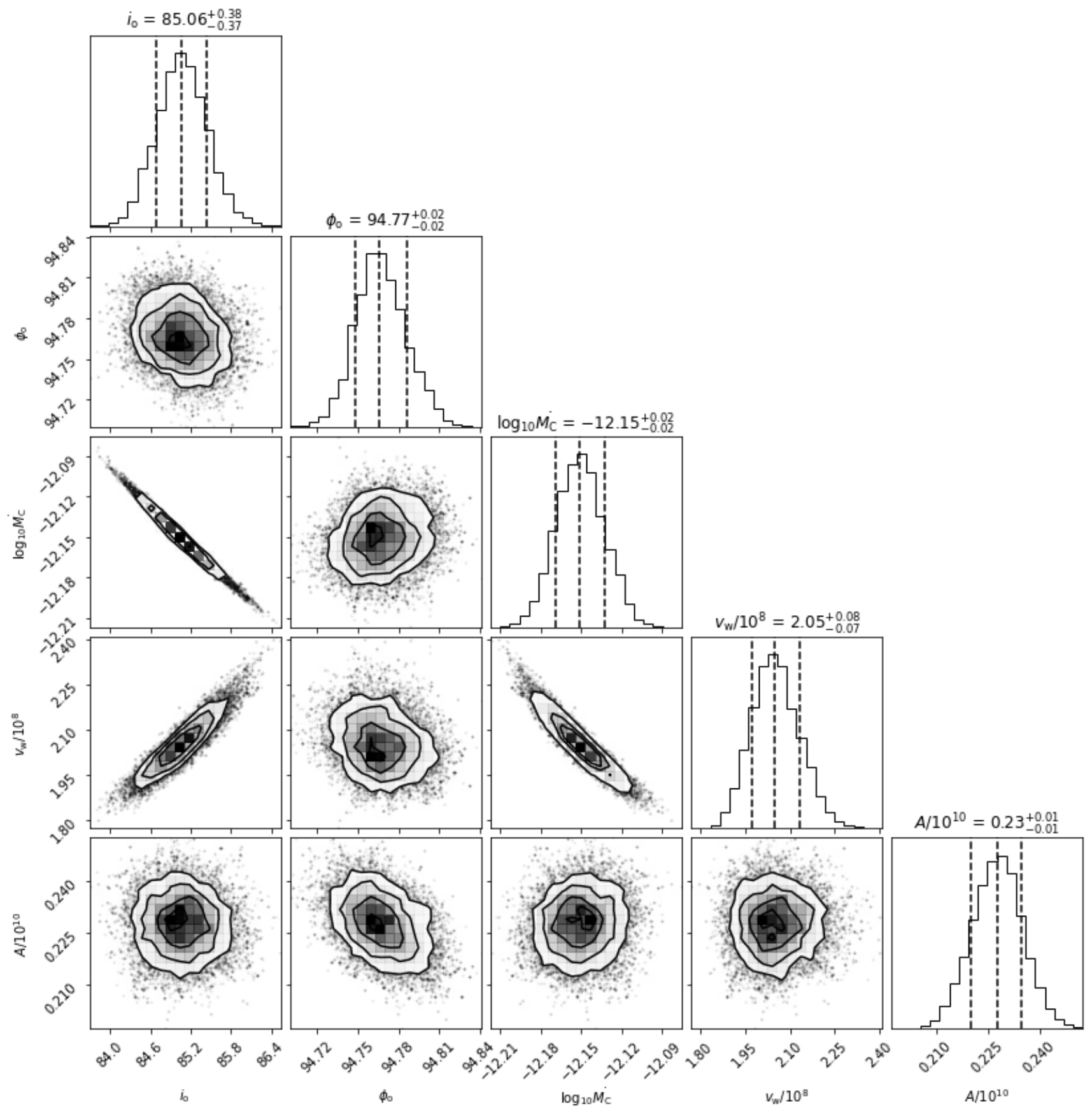}
    \caption{\textit{Left:} Normalized flux density and DM variations of PSR B1957$+$20 during the eclipse phase. The solid lines give the model fittings for the observational data. \textit{Right:} Corner plot showing the posterior distributions of the model parameters.}
    \label{Fig.2}
\end{figure*}
   
PSR B1957$+$20 is the first MSP discovered in a binary with a low-mass companion star, and it is classified to be a BW system. The period and eccentricity of the binary orbit is $P_{\mathrm{orb}} \sim 9.2 \, \mathrm{hours}$ and $e < 4.0 \times 10^{-5}$ \citep{Arzoumanian_1994}. 
PSR B1957+20 has a spin period of 1.607 ms and a spin-down luminosity of $L_{\mathrm{sd}} \sim 1.6 \times 10^{35} \, \mathrm{erg \, s^{-1}}$ \citep{Fruchter_1988}. During the eclipse phase, 
the flux density of PSR B1957+20 also decreases to nearly zero, which makes it unable to obtain the value of DM. In the left panel of Figure \ref{Fig.2}, we present the observational data centered at $\nu \sim 1.25 \, \mathrm{GHz}$ on 02 Sep. 2022, which covers the whole eclipsing period and lasts 14340 s, where the flux density is normalized by the pre- and post-eclipse emission (Wang et al. in prep). Solid lines represent fittings of the data, and the goodness of fitting is evaluated with the MCMC method. The observational constraints on the model parameters are shown in the right panel of Figure \ref{Fig.2} and the corresponding parameter values with $1-\sigma$ errors are listed in Table \ref{T1}.

The obtained inclination angle of $\sim {85.06^{\circ}}^{+0.38^{\circ}}_{-0.37^{\circ}}$ indicates the orbit of PSR B1957$+$20 is nearly viewed edge-on, which is consistent with the result ($ i_{\mathrm{o}} \sim 85^{\circ}$) given by \cite{Johnson_2014} through their MSP light curves simulation model. In comparison, \cite{Reynolds_2007} derived a relatively smaller inclination of $i_{\mathrm{o}} = 65^{\circ} \pm 2^{\circ}$ from the photometric data of the William Herschel Telescope and Hubble Space Telescope. 
Alternatively, \cite{Kandel_2021} constrained the inclination to be $i_{\mathrm{o}} \sim 75.8^{\circ} \pm 5.9^{\circ}$ by using the X-ray emission arising from the stellar wind interactions.

The mass-loss rate of the companion of PSR B1957$+$20 is constrained to $\dot{M_{\mathrm{C}}} \sim 10^{-12.15} \, \mathrm{M_{\odot} \, yr^{-1}}$, which is also well consistent with the result (ie., $\dot{M_\mathrm{C}} \sim 10^{-12} \, \mathrm{M_{\odot} \, yr^{-1}}$) of \cite{Polzin_2020} who analyzed the 149 MHz observations with an inclination of $65^{\circ}$ \citep{Reynolds_2007}. It is considered that the mass loss of the companion star is caused by the evaporation due to the irradiation from the MSP and, thus, its rate can be estimated by
\begin{equation}
    \dot{M}_{\mathrm{C}} = - \frac{f}{2 v^{2}_{\mathrm{w}}} L_{\mathrm{sd}} \left( \frac{r_{\star}}{a_{\mathrm{orb}}} \right)^{2} \label{12}
\end{equation}
where $a_{\mathrm{orb}}$ is the orbital separation of the binary and $f$ is the evaporation efficiency. Then, by using $\dot{M_{\mathrm{C}}} \sim 10^{-12.13} \, \mathrm{M_{\odot} \, yr^{-1}}$, we can obtain $f=2.35\times10^{-3}$ for PSR B1957$+$20. Furthermore, we can in principle explore a clue from the mass-loss rate to determine the evolutionary stage of the companion star \citep{Chen_2013}. 

\subsection{PSR J2051$-$0827} \label{subsec:PSR J2051-0827}

\begin{figure*}[h!]
    \centering
    \includegraphics[scale = 0.4]{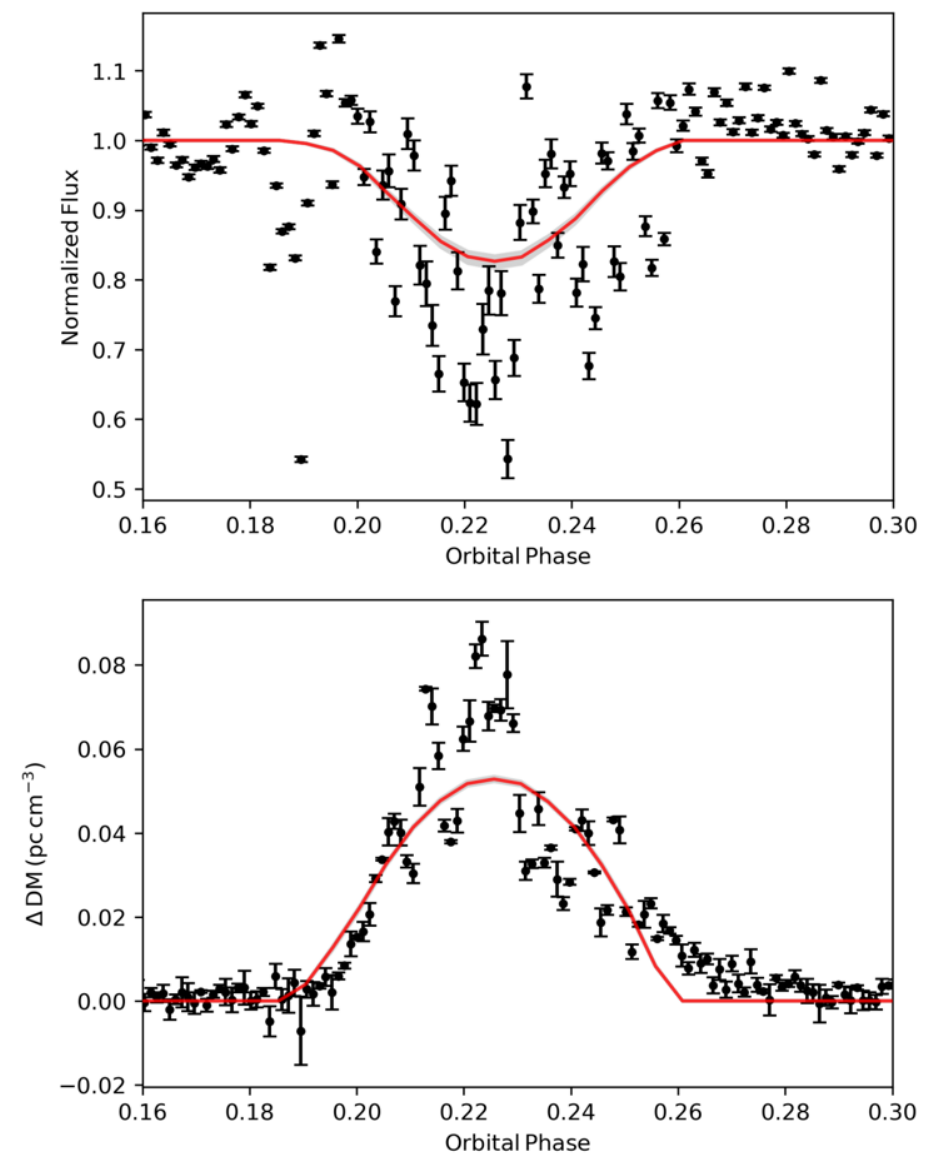} \includegraphics[scale = 0.25]{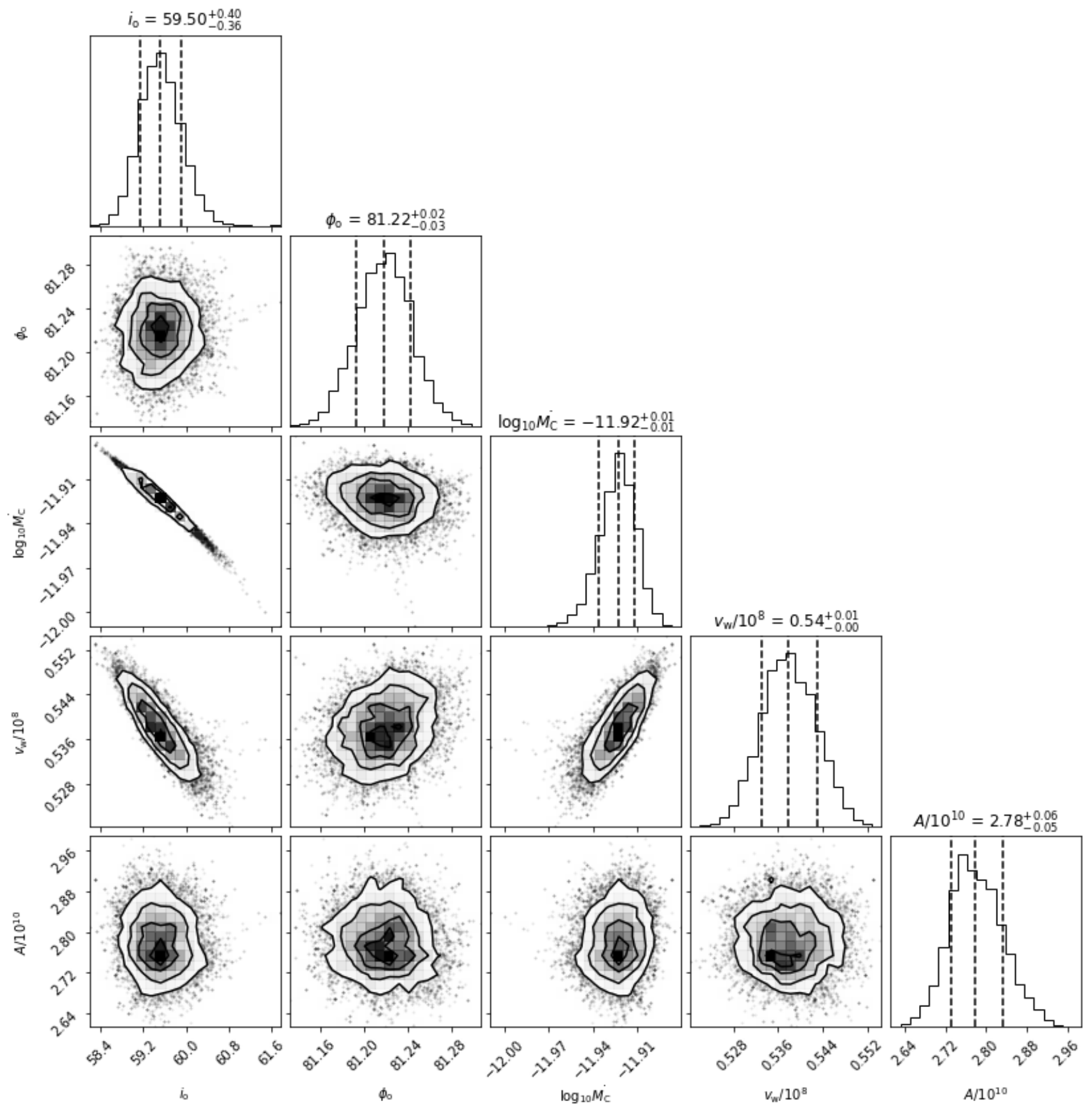}
    \caption{The same to Figure \ref{Fig.2} but for PSR J2051$-$0827.}
    \label{Fig.3}
\end{figure*}
   
PSR J2051$-$0827 is the second eclipsing MSP discovered after PSR B1957$+$20. The MSP is in a 2.38-hour orbit around the companion star with an eccentricity of $3 \times 10^{-4}$. The pulsar has a spin period of $4.508$ ms and a spin-down luminosity of $6 \times 10^{33}  \mathrm{erg \, s^{-1}}$. Radio observations at low frequency ($\nu \leq 0.6$ GHz) show eclipse features covering $10 \%$ of the orbital period \citep{Stappers_1996-asps}. 
The FAST observations of PSR J2051$-$0827 centered at $\nu \sim 1.26$ GHz was carried out on 14th Jan. 2022 with a duration of 8940s, covering the entire binary orbit \citep{Wang_2023}. 
The observational data are displayed in the left panel of Figure \ref{Fig.3}, which shows the radio emission of the pulsar was not completely obscured during the eclipse.
Furthermore, in comparison with the previous source, the data of PSR B1957$+$20 has a much larger scatter, which makes it impossible to obtain a good fitting with a simple model. A possible explanation for the large fluctuation of the data is that the stellar wind in this source is very unstable and in-homogeneous.

In any case, by according to a rough fitting to the data of PSR J2051$-$0827, we can still get a constraint on the inclination angle as ${59.50^{\circ}}^{+0.40^{\circ}}_{-0.36^{\circ}}$. In the previous work of \cite{Dhillon_2022}, this angle was found to be ${55.9^{\circ}}^{+4.8^{\circ}}_{-4.1^{\circ}}$ by fitting the symmetrical HiPERCAM light curves with a direct-heating model of the pulsar to the companion star. Meanwhile, \cite{Stappers_2001} estimated it to be $\sim 40 ^{\circ}$ by using the asymmetry optical light curves of the companion, where the gravitational distortion of the companion and the irradiation of the pulsar wind are taken into account. The mass-loss rate of the companion $\dot{M_{\mathrm{C}}} \sim 10^{-11.92} \, \mathrm{M_{\odot} \, yr^{-1}}$ is also found to be close to the previous estimate by \cite{Polzin_2019}, which determines an evaporation efficiency of $3.33\times10^{-2}$. 

\subsection{PSR J2055$+$3829} \label{subsec:PSR J2055+3829}

\begin{figure*}[h!]
    \centering
    \includegraphics[scale = 0.4]{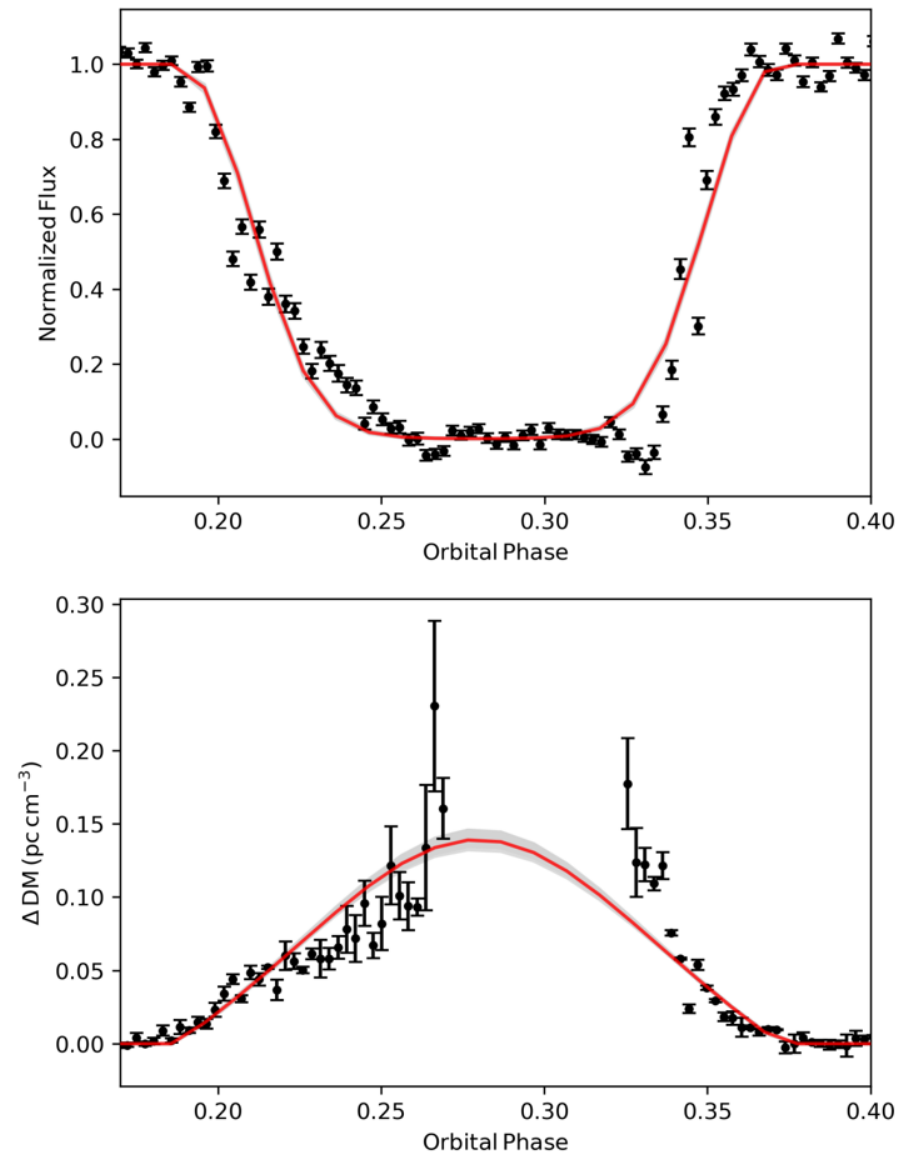} \includegraphics[scale = 0.25]{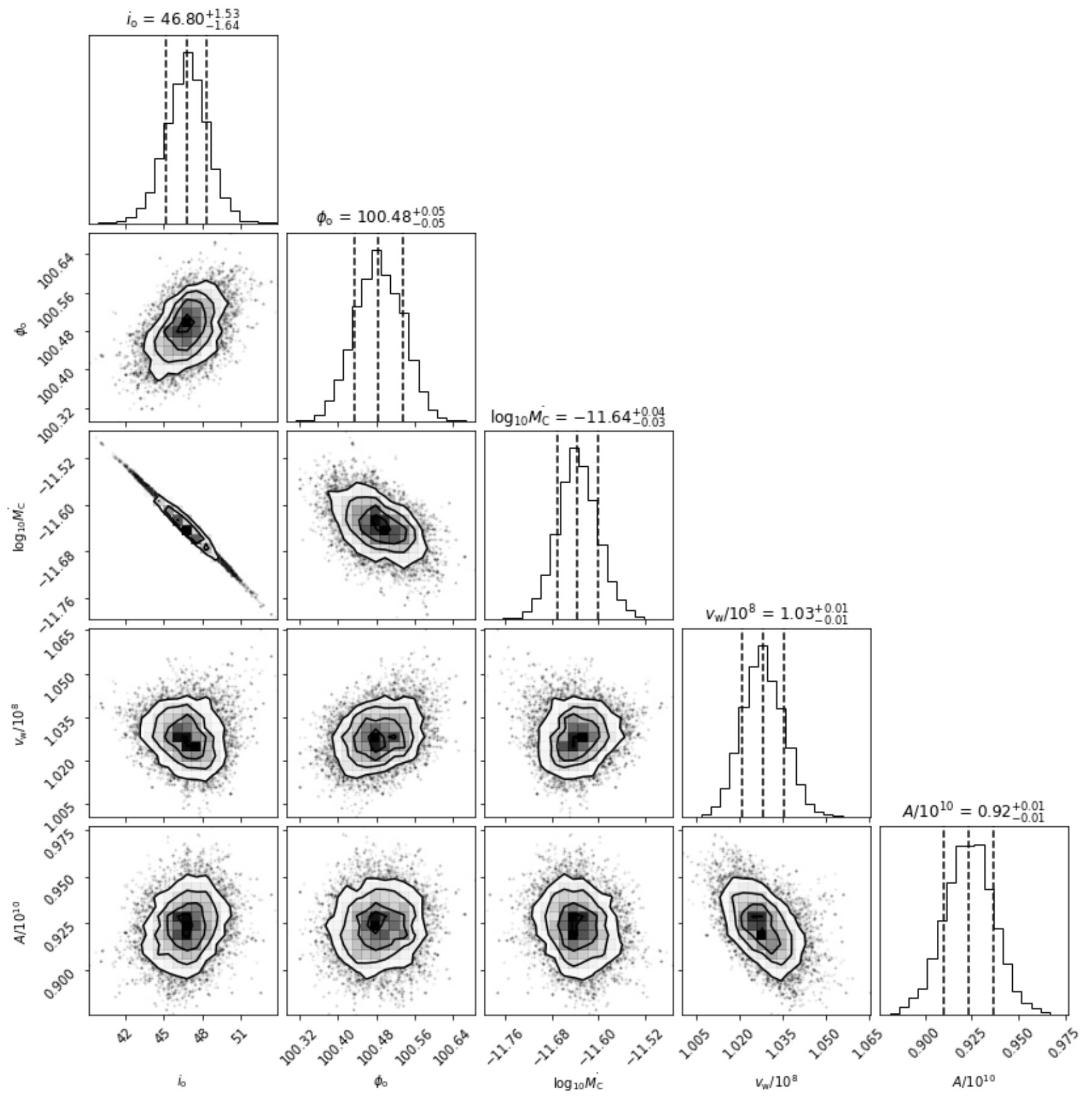}
    \caption{The same to Figure \ref{Fig.2} but for PSR J2055$+$3829.}
    \label{Fig.4}
\end{figure*}
   
PSR J2055$+$3829 was first discovered by the SPAN512 survey conducted with the Nan{\c c}ay Radio Telescope \citep{Octau_2018}, 
which has a spin period of 2.08 ms and a spin-down luminosity of $L_{\mathrm{sd}} \sim 4.3 \times 10^{33} \, \mathrm{erg \, s^{-1}}$ \citep{Guillemot_2019}. As another BW pulsar, PSR J2055$+$3829 is in a binary with a very low mass companion of a tight 3.1-hour orbit. Wang et al.(in prep) conducted observations on PSR J2055$+$3829 for 11640 s centred at 1.25 GHz on 30th Aug. 2021, and found a clear eclipse with a duration of 1467$\pm$12 s, which is about 13\% of the orbital period. The observational data are shown in the left panel of Figure \ref{Fig.4}, in comparison with the modelling curves.

The fitting results indicate a relatively moderate inclination angle of $i_{\mathrm{o}} \sim {46.80^{\circ}}^{+1.53^{\circ}}_{-1.64^{\circ}}$, which is between the edge-on orbit ($i \sim 90^{\circ}$) assumed by \cite{Guillemot_2019} and the inclination of $26^{\circ}$ estimated from the mass function with an upper limit on the companion mass. The mass loss rate and the velocity of the companion wind are constrained to be $\dot{M_{\mathrm{C}}} \sim 10^{-11.64} \,  \mathrm{M_{\odot} \, yr^{-1}}$ and 1.03 $\times$ $10^{8} \, \mathrm{cm \, s^{-1}}$, respectively, indicating an evaporation efficiency of $f=4.34\times10^{-2}$.

\section{Discussion and implications}
\label{sect:discussion}

\subsection{RM measurements and magnetic fields}

\begin{figure*}[h!]
    \centering
    \includegraphics[scale = 0.45]{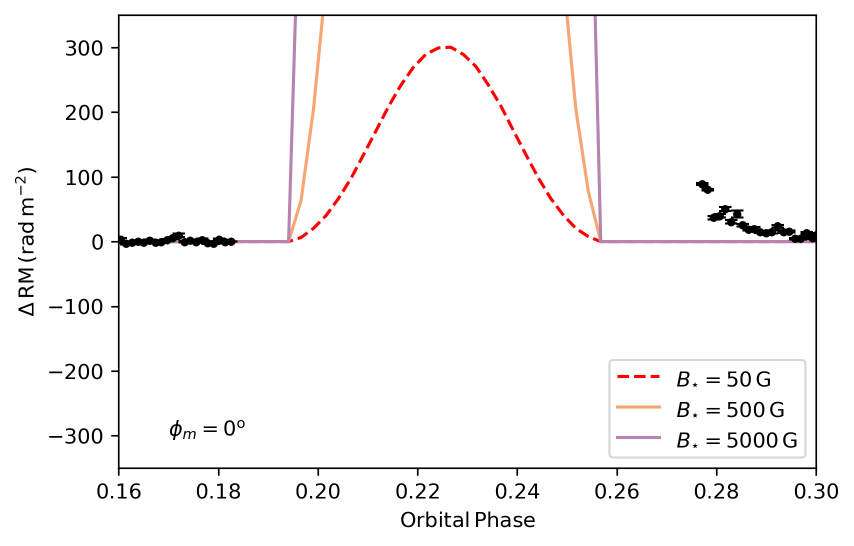} \includegraphics[scale = 0.45]{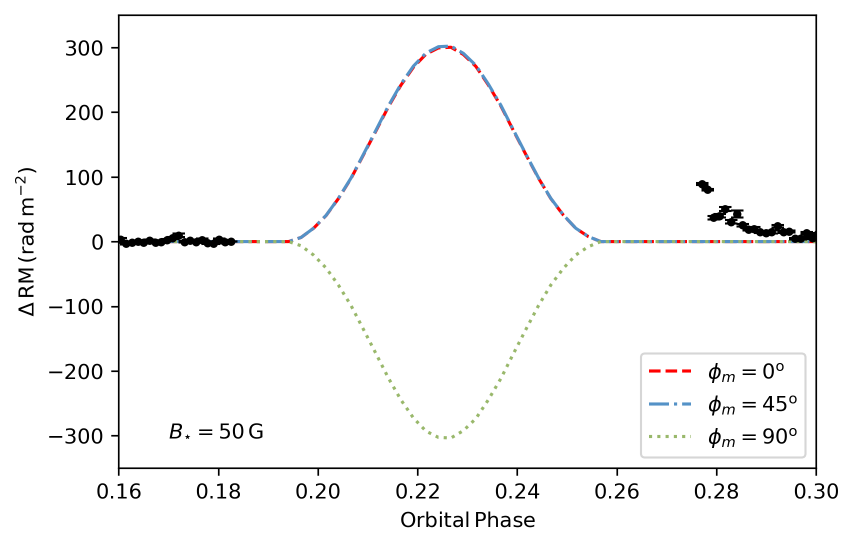}
    \caption{A comparison of the model-predicted RM variation of PSR J2051$-$0827 during the eclipse phase with the observational data (solid circles), where a typical magnetic field structure is adopted as presented in Equation (\ref{Bstr}). Different values for the parameter $B_{\star}$ and the inclination of the magnetic axis $\phi_{\rm m}$ are taken as labeled in the legend, while the other model parameters are taken as listed in Table \ref{T1}. The observational data are taken from \cite{Wang_2023}.}
   \label{Fig.5}
\end{figure*}

For PSR J2051$-$0827, \cite{Polzin_2019} utilized the linearly and circularly polarized flux to investigate the possible presence of magnetic fields through the measurements of rotation measure (RM), Faraday delay, and depolarization.
As a result, the average magnetic field parallel to the LOS was constrained to be within the range of $10^{-4} \, \mathrm{G} \leq B_{\parallel} \leq 10^{2}$ G. Very recently, \cite{Wang_2023} detected a regular decrease of the RM of PSR J2051$-$0827 in the egress of the eclipse. i.e., the RM gradually change back to normal when the LOS moves away from the eclipse medium, as shown by the data points in Figure \ref{Fig.5}. This measurement provided direct evidence for the existence of magnetic fields in the eclipse medium.  

In theory, based on the obtained constraints on the orbital and companion wind properties, we can in principle predict the possible variation of RM of the spider pulsars by
\begin{align}
    \Delta \mathrm{RM} &= \frac{q_{\mathrm{e}}^3}{2 \pi m_{\mathrm{e}}^2 c^4} \int_{l_{\mathrm{p,obs}}}^{\infty} n_{\mathrm{e}} \textbf{\emph{B}}_{\mathrm{w}} \, d \textbf{\emph{l}}, \label{9}
\end{align}
by adopting a specific description of the magnetic field in the stellar wind ($\textbf{\emph{B}}_{\mathrm{w}}$). Typically, we can express the magnetic field as a combination of the radial and toroidal components as \citep{Bosch-Ramon_2011}:
\begin{equation}
\textbf{\emph{B}}_{\mathrm{w}} (r, \phi) = B_{\mathrm{r}} \textbf{\emph{e}}_{\mathrm{r}} + B_{\phi} \textbf{\emph{e}}_{\phi}\label{Bstr}
\end{equation}
with 
\begin{align}
    B_{r} &\sim B_{\star} (r_{\star} / r)^2, \label{4}\\
    B_{\phi} &\sim B_{\star} (v_{\mathrm{w}, \phi} / v_{\mathrm{w}}) (r_{\star} / r), \label{5}
\end{align}
where $B_{\mathrm{\star}}$ is the magnetic field strength at $r_{\mathrm{\star}}$ and the surface rotation velocity of the companion star is assumed to be $v_{\mathrm{w}, \phi} \sim 0.1 v_{\mathrm{w}} (r_{\star} / r)$. Finally, it needs to be pointed out that the symmetric axis of the above magnetic field could be not parallel to the angular momentum of the binary, but probably has an inclination angle of $\phi_{\rm m}$ relative to the normal of the orbital plane, which would also influence the RM variation. Here the binary is assumed to be tidally locked.

By using the above formulae, we plot the theoretical RM variation curves in Fig. \ref{Fig.5} for different magnetic field strengths and different magnetic inclinations, where the orbital and wind parameters are taken as listed in Table \ref{T1} for PSR J2051$-$0827. However, by comparing the theoretical curves with the observational data, it is unfortunately found that the model always cannot explain the data. An obvious difference is that the data is delayed by a time of about 0.015\% of the orbital period relative to the model prediction. This indicates that it is necessary to consider a more complicated structure for the magnetic field, in particular, around the bow shock \citep[e.g.,][]{Phinney_1988}.
Due to the fact that the magnetic field near the bow shock can be significantly compressed by the shock and, furthermore, the distribution of the magnetic field may not completely trace the bow shock. In particular, the LOS could be scanned by the tail of the bow shock for a long period because the shock tail can be deflected due to the orbital motion. In this tail region, the magnetic field could be very high so that a significantly high RM contribution can be expected, whereas the corresponding radio absorption and DM contribution are negligible.
Additionally, the orbital motion of the companion would not only have an influence on the shock tail but would also lead the symmetric axis of the bow shock to deflect from the connecting line of the binary. The deflection angle of the bow shock can in principle be estimated by $\xi \sim \mathrm{arctan} (v_{\mathrm{C}} / v_{\mathrm{w}})$, where $v_{\mathrm{C}}$ is the orbital velocity of the companion \citep{Zabalza_2013}. For PSR J2051$-$0827, we can get $\xi \sim 47.5^{\circ}$, which is much larger than the other two BW binaries.
Therefore, the consideration of these complications about the magnetic field configuration around the bow shock is definitely needed in future modeling works as more RM data can be obtained.

\subsection{The companion surface temperature and pulsar masses}

As the radius of the companion star of PSR J2051$-$0827 had been measured to be $r_{\star} = 0.139^{+0.011}_{-0.015} \, R_{\odot}$ by \cite{Dhillon_2022}, we can roughly derive its surface temperature to be $T_{\star}=6093^{+131}_{-109} \, \mathrm{K}$ from the obtained parameter value of $A = 2.78^{+0.06}_{-0.05} \times 10^{10} \, \mathrm{K \, cm^{2/3}}$. This temperature is higher than the one given by \cite{Stappers_1996-apjl-b} as $T_{\star} = 4000 - 4700 \, \mathrm{K}$, which was estimated from the observed $R-I$ color according to the cool M$-$type dwarf relations of Bessell.
For PSR J2055$+$3829, if we use the Roche lobe radius $R_{\mathrm{L}} \sim 0.14~R_{\odot}$ to approximate the companion radius, which is reasonable for an appropriate mass ratio between the pulsar and the companion \citep{Guillemot_2019}, we can estimate the surface temperature of the companion to be $2016^{+22}_{-22} \, \mathrm{K}$ from the parameter value $A = 2.78^{+0.06}_{-0.05} \times 10^{10} \, \mathrm{K \, cm^{2/3}}$. However, for PSR B1957$+$20, the combination of the obtained parameter $A=0.23^{+0.01}_{-0.01} \times 10^{10} \, \mathrm{K \, cm^{2/3}}$ with the possible radius $r_{\star} \simeq 0.25 R_{\odot}$ of the companion \citep{van-Kerkwijk_2011} would lead to a temperature as low as $T_{\star} \simeq 343^{+15}_{-15} \, \mathrm{K}$, which is much lower than that found by \cite{Reynolds_2007} and is probably unrealistic. This indicates that the value of $\beta=2/3$ may not be always appropriate for the companions or Equation (\ref{Tr}) for a constant index is not strictly correct, especially for the wind very close to the stellar surface. 
In addition, we would like to emphasize again that the temperature of the companion should in fact be aisotropic due to the irradiation of pulsar wind and the temperature obtained from the fitting of the radio flux and DM is just an effective value, which might be regarded as an average temperature.

\begin{figure}[h!]
    \centering
    \includegraphics[scale = 0.6]{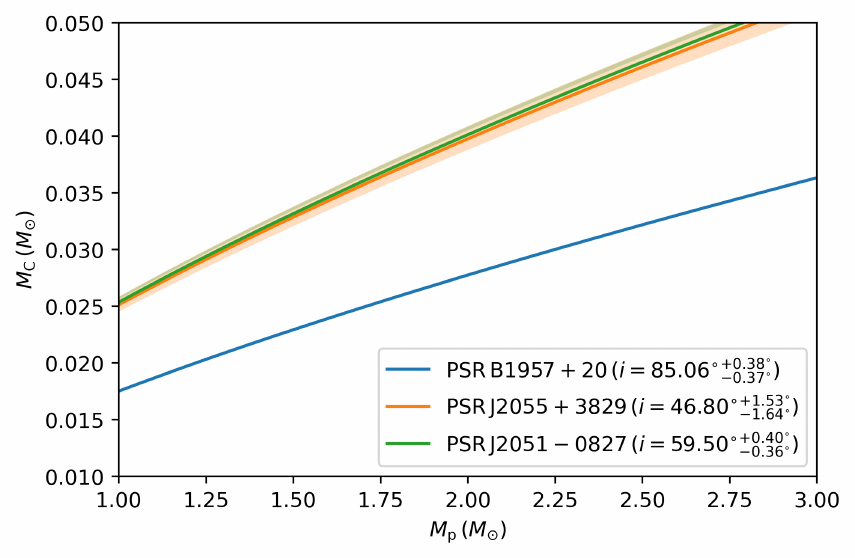}
    \caption{The relationship between the masses of PSR B1957$+$20, PSR J2055$+$3829, and PSR J2051$-$0827 and their companions.}
    \label{Fig.6}
\end{figure}

In any case, the determination of the companion surface temperature would in principle help us to calculate their emission luminosity and, furthermore, to estimate the masses of the companions by according to an empirical mass-luminosity relationship. Meanwhile, the determination of the inclination angle of the BW binaries enables us to fix the mass function of the binaries as 
\begin{equation}
    \left( \frac{2 \pi}{P_{\mathrm{orb}}} \right)^2 \frac{(a_{\mathrm{p}} \mathrm{sin} \, i_{\mathrm{o}})^3}{G} = \frac{(M_{\mathrm{C}} \mathrm{sin} \, i_{\mathrm{o}})^3}{(M_{\mathrm{p}}  + M_{\mathrm{c}})^2}, \label{EQ4.3}
\end{equation}
where $M_{\rm p}$ and $M_{\rm C}$ are the masses of the pulsar and companion, respectively. To be specific, the $M_{\rm p}-M_{\rm C}$ relationships of the three FAST BW binaries are displayed in Fig. \ref{Fig.6}. Then, if the companion masses can indeed be obtained from the temperature measurement, it is expected that the pulsar masses would be finally constrained. 

\section{Summary and conclusion}

The study of pulsar eclipse processes occurring in BW systems can provide a more accurate perception of binary orbits and their companions. 
It is proposed that the radio emission from the pulsar is absorbed by the outflows of its companion. 
The interaction of the pulsar wind with the stellar wind forms a bow shock, and the stellar outflows in the shock-wrapped companion side cause the radio eclipse when the star moves around inferior conjunctions. 

We establish an eclipse model of wind interaction scenario and fit the radio flux density and DM variations of three BW pulsars detected by FAST, namely PSR B1957$+$20, PSR J2055$+$3829, and PSR J2051$-$0827. 
We constrain some of the orbital parameters as well as the wind parameters of companion stars under the eclipse model.
For PSR B1957$+$20, we found that the binary has a relatively high orbital inclination ($i_{\mathrm{o}} = {85.06^{\circ}}^{+0.38^{\circ}}_{-0.37^{\circ}}$) and a true anomaly of the observer of $\phi_{\mathrm{o}} = {94.77^{\circ}}^{+0.02^{\circ}}_{-0.02^{\circ}}$.The mass loss rate and the wind velocity of the low-mass companion are $\dot{M_{\mathrm{C}}} \sim 10^{-12.15} \, \mathrm{M_{\odot} \, yr^{-1}}$ and $v_{\mathrm{w}} \sim 2.05 \times 10^{8} \, \mathrm{cm \, s^{-1}}$.
While for PSR J2051$-$0827, we obtain the orbital parameters such as its orbital inclination of $i_{\mathrm{o}} = {59.50^{\circ}}^{+0.40^{\circ}}_{-0.36^{\circ}}$ and the true anomaly of the observer as $\phi_{\mathrm{o}} = {81.22^{\circ}}^{+0.02^{\circ}}_{-0.03^{\circ}}$, the companion's mass loss rate and wind velocity are constrained to be $\dot{M_{\mathrm{C}}} \sim 10^{-11.92} \, \mathrm{M_{\odot} \, yr^{-1}}$ and $v_{\mathrm{w}} \sim 0.54 \times 10^{8} \, \mathrm{cm \, s^{-1}}$, respectively.
Similarly, for PSR J2055$+$3829, we use the eclipse model to constrain the orbital inclination as $i_{\mathrm{o}} = {46.80^{\circ}}^{+1.53^{\circ}}_{-1.64^{\circ}}$ and the true anomaly of the observer as $\phi_{\mathrm{o}} = {100.48^{\circ}}^{+0.05^{\circ}}_{-0.05^{\circ}}$, and the mass loss rate and the wind velocity of the companion are constrained to be $\dot{M_{\mathrm{C}}} \sim 10^{-11.64} \, \mathrm{M_{\odot} \, yr^{-1}}$ and $v_{\mathrm{w}} \sim 1.03 \times 10^{8} \, \mathrm{cm \, s^{-1}}$, respectively.
The above fitting results indicate that our eclipse model can be used to constrain the orbital parameters as well as the wind parameters of the companions in BW binary systems.

In addition, our eclipse model still needs further refinement, as for the evidence of the magnetic field in the RM variation detected in the late stage of the eclipse. We suggest that it is not dominated by the magnetic field of the companion star but by the interaction between the magnetic field of the pulsar wind and the bow shock, which relies on a more accurate description of the magnetic field around the bow shock.
Furthermore, if the mass and radius of the companion can be given directly by optical observations, we can also make an estimate of the mass of the pulsar in the BW systems by eclipse.

\begin{acknowledgements}
We thank the referee for useful comments on the manuscript.
This work is supported by the National SKA program of China (2020SKA0120300), the National Key R\&D Program of China (2021YFA0718500), the National Natural Science Foundation of China (grant Nos. 11833003, 12033001), the China Postdoctoral Science Foundation
(No. 2023T160410), and the Opening Foundation of Xinjiang Key Laboratory (No. 2021D04016).
\end{acknowledgements}

\bibliography{RAA-2023-0245}{}
\bibliographystyle{raa}

\label{lastpage}

\end{document}